\documentstyle[11pt,paspconf,psfig]{article}

\begin{document}

\title{Fast X-ray Variability of PKS 2155--304: A Cross Correlation
Analysis}

\author{A.\ Treves}
\affil{Universit\'a di Como, Via Lucini 3, I-22100 Como, Italy}
\author{Y.H. \ Zhang, A. \ Celotti}
\affil{SISSA/ISAS, Via Beirut 2-4, I-34014 Trieste, Italy}
\author{F. \ Tavecchio, L.\ Maraschi}
\affil{Osservatorio Astronomico di Brera, Via Brera 28, I-20121 Milano, 
Italy} 
\author{G. \ Ghisellini, G. \ Tagliaferri}
\affil{Osservatorio Astronomico di Brera, Via Bianchi 46, I-22055
Merate (Lecco), Italy}
\author{L.\ Chiappetti}
\affil{IFCTR, C.N.R., Via Bassini 15, I-20133 Milano, Italy}
\author{E. Pian}
\affil{ITESRE, C.N.R., Via Gobetti 101, I-40129, Bologna, Italy}
\author{C. M. \ Urry}
\affil{STScI, 3700 San Martin Drive, Baltimore MD 21218, USA}


\keywords{PKS 2155-304}

\section{Introduction}

PKS 2155--304 is one of the prototypes of BL Lac objects. Its spectral
energy distribution is known from the radio to the TeV band (e.g.,
Chiappetti et al. 1998; Chadwick et al. 1998). 
The emitted luminosity in the $\nu F(\nu)$
representation has two peaks, one in the soft X-rays and the other in 
the TeV region. This spectral
shape, which is typical of High Frequency peaked (X-ray selected) BL
Lacs, may be interpreted within the synchrotron self-Compton model,
which attributes the first peak to synchrotron radiation of electrons
which also produce the second peak via Compton scattering of the same
synchrotron photons. The emission is relativistically boosted in the
observer direction.
A way of testing the model and characterizing its geometrical and
kinematical properties is through the study of correlated variability
in different bands. Because of its brightness, PKS 2155--304 is one of
the few objects where such a study can be performed. In 1994,
simultaneous observations with ASCA, EUVE and IUE (Urry et al. 1997)
were carried out demonstrating the presence of a 1 day time lag
between UV (2000 \AA) and soft X-rays (100 \AA), the latter lagging in
turn by 1 day with respect to the 2-10 keV X-rays. A lag of 1 h was
also found between the 0.5-1 and 2-10 keV photons (Makino et
al. 1996).  The behaviour of the source was clearly different from
that observed in 1991 (Edelson et al. 1995), when the comparison of
IUE and ROSAT data demonstrated that the lag between UV and 1-4 keV
photons was of the order of 2 hr.
Here we report on the search and accurate study of lags in two long
X-ray observations performed with the $Beppo$SAX satellite. The first
one was accomplished during the performance verification (PV) phase of the
satellite in Nov. 1996 (Giommi et al. 1998), the second one in
Nov. 1997, in correspondence of a high activity phase of the source in
gamma-rays (Maraschi et al. 1998; Chiappetti et al. 1998). $Beppo$SAX
covers a broad energy interval extending from 0.1 to 200 keV and it is
therefore apt for searching for lags or leads within the X-ray
band. The cross correlation technique requires of course that the
light curves are binned on time intervals smaller than the searched
lags. We are therefore practically limited to the use of the 0.1-10
keV band, which is covered by the low energy concentrator spectrometer
(LECS, 0.1-10 keV) and the medium energy concentrator spectrometers
(MECS 2-10 keV). For brevity here we focus on the results relative to
the $Beppo$SAX PV phase observations only.  In a forthcoming paper
(Zhang et al., in preparation) we will also report on a similar
analysis applied to the Nov. 1997 $Beppo$SAX and 1994 ASCA data
retrieved from the archives.

\section{Light Curves}

The procedure applied for producing the $Beppo$SAX light curves is
discussed in detail in Chiappetti et al. (1998). Fig. 1 shows the LECS
and MECS count rates in the 0.1-1.5 keV and 3.5-10 keV energy bands
and the hardness ratio (HR) between them. The HR shape is similar to
that of the light curves, a direct indication of the lead of higher
energies.


\begin{center}
\begin{figure}
\vspace{0.cm}
\begin{tabular}{cc}
\psfig{figure=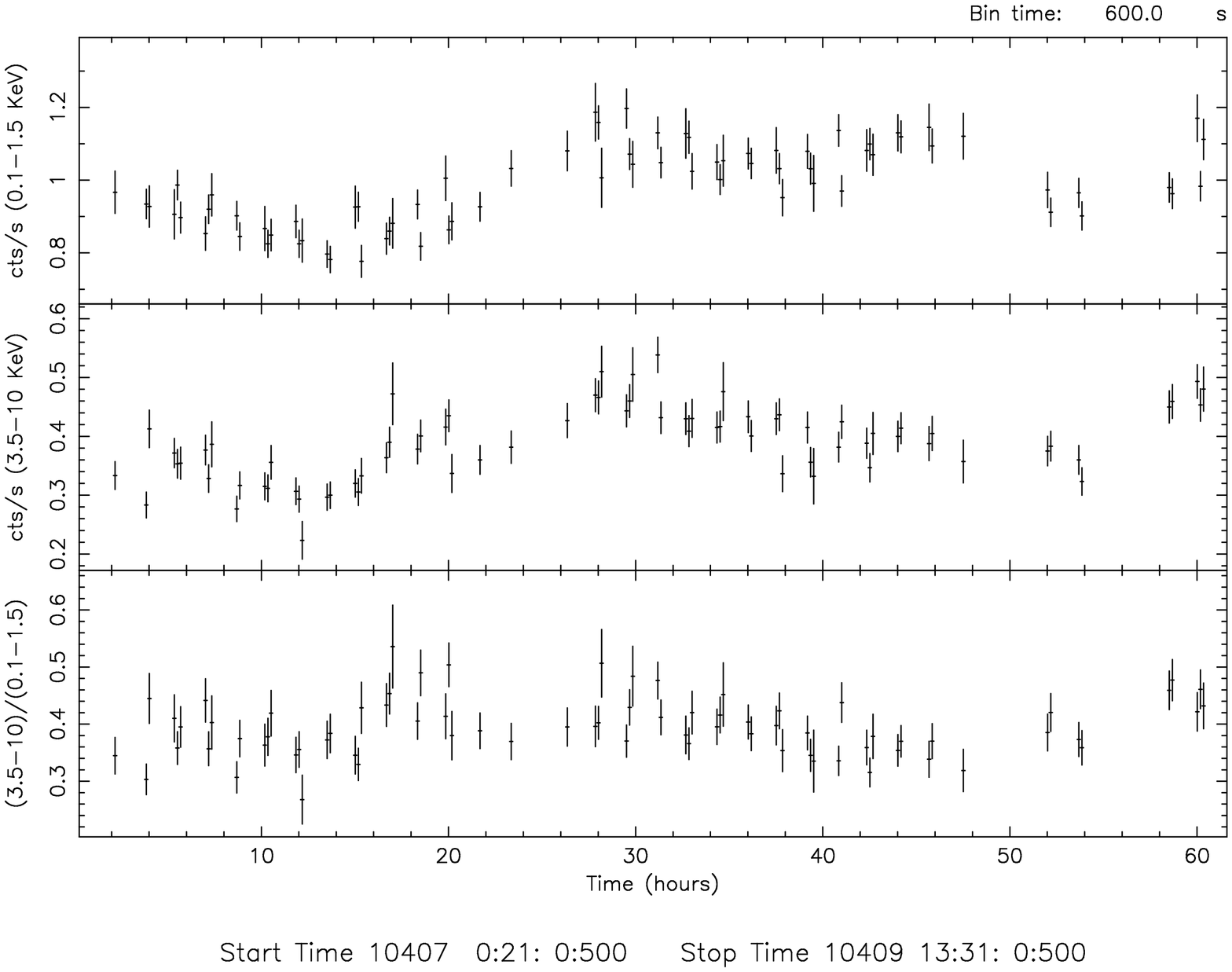,angle=270,width=6cm,height=6cm,clip=}
\psfig{figure=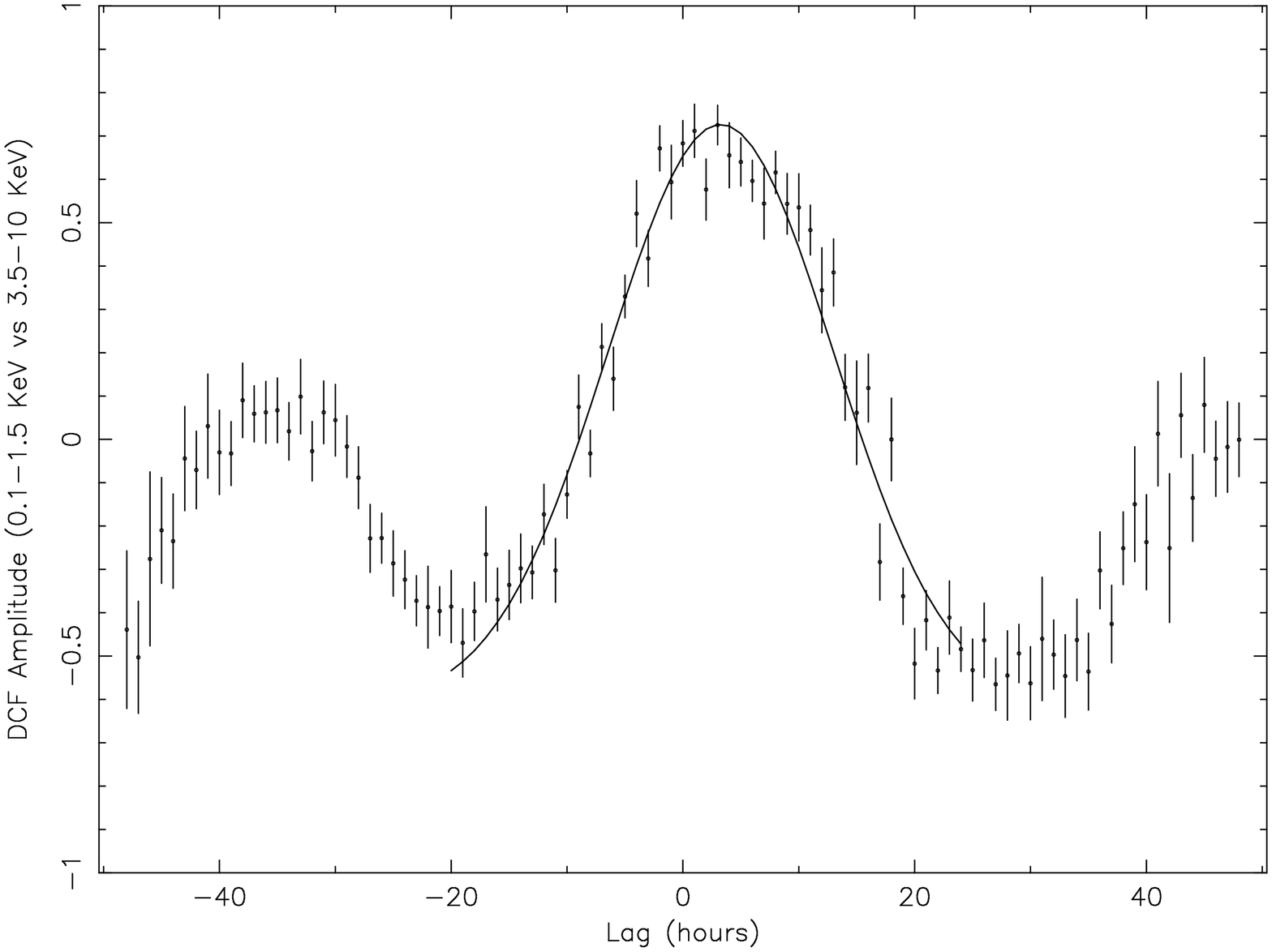,angle=270,width=6cm,height=6cm,clip=}
\end{tabular}
\caption{Left panel: the light curves and HR 
of PKS~2155--304 in November 1996. Right panel: DCF cross correlation with
gaussian fit.}
\label{fig-1b} \end{figure}
\end{center}

\begin{center}
\begin{figure}
\vspace{0.cm}
\begin{tabular}{cc}
\psfig{figure=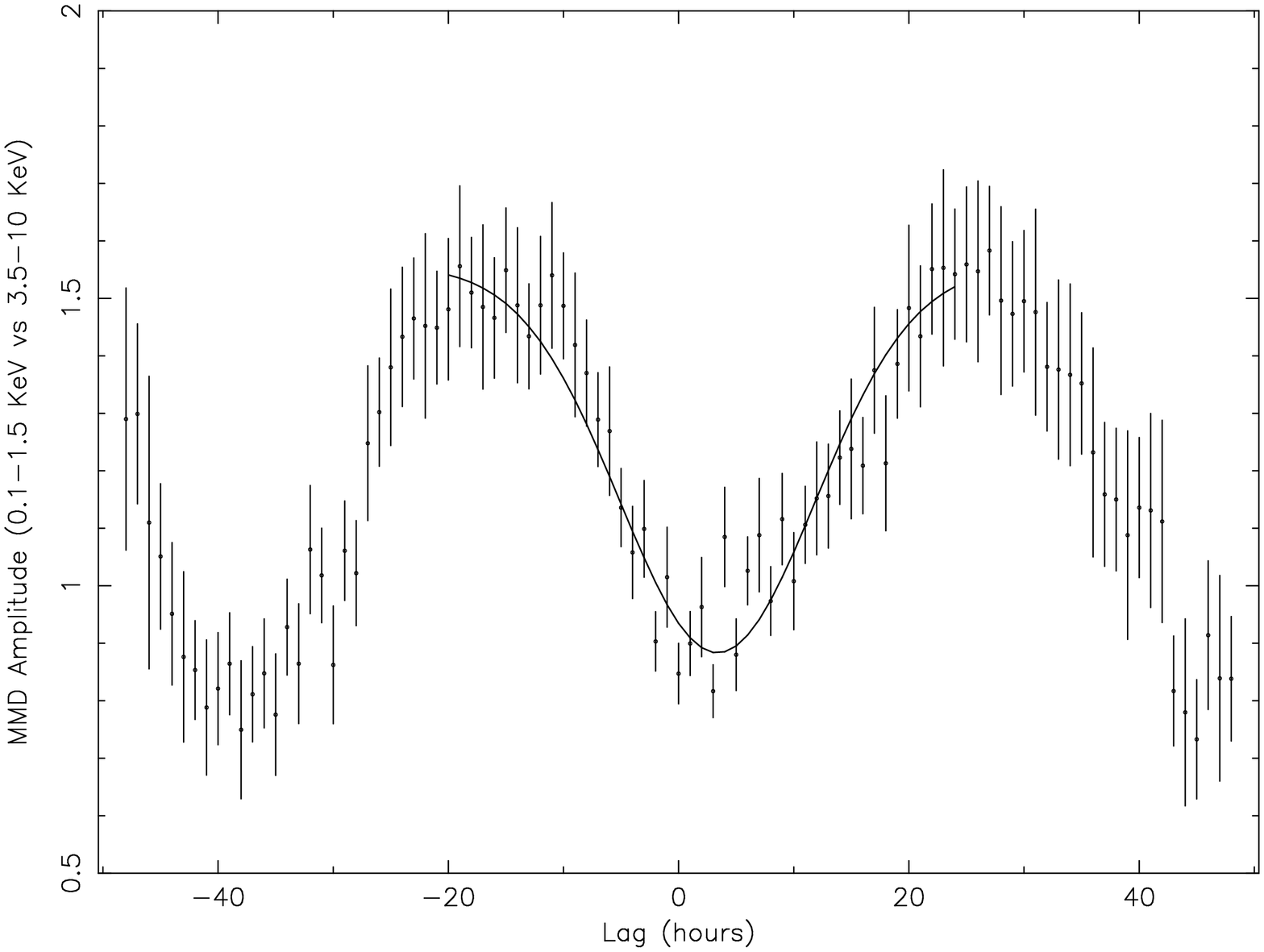,angle=270,width=6cm,height=6cm,clip=}
&\psfig{figure=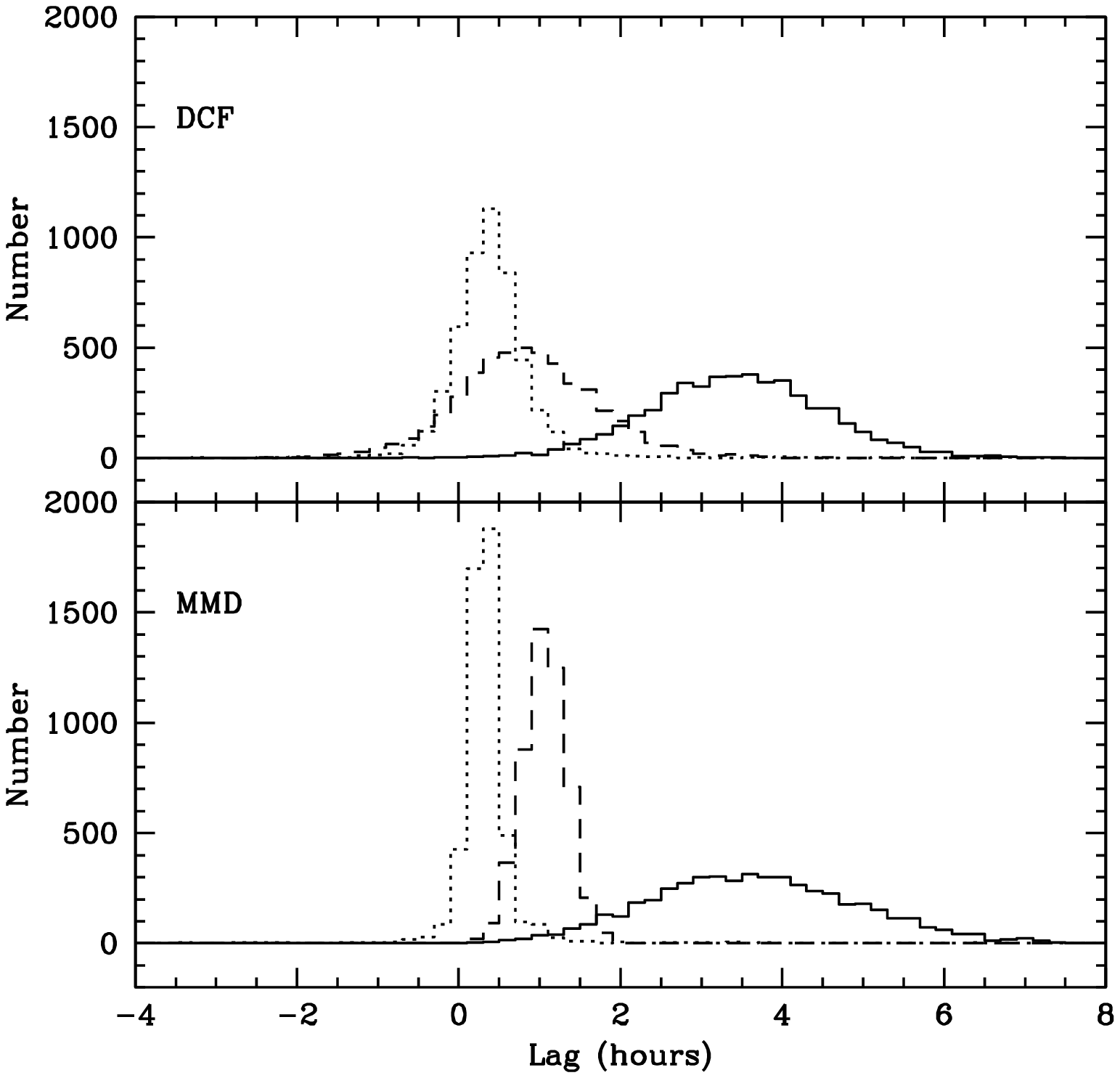,width=6.5cm,height=6.5cm}
\end{tabular}
\caption{Left panel: MMD with gaussian fit. 
Right panel: FR/RSS Simulations. The solid line refers to the 
SAX 1996, the dotted line
to the SAX 1997 and the dashed line to the ASCA 1994 data.}
\end{figure}
\end{center}

\section{Cross Correlation Analysis}

\subsection{Discrete Correlation Function (DCF)}

The DCF method is described in Edelson \& Krolik (1988).
The DCF results are reported
in Fig. 1; the chosen DCF binning size was 3600 sec
and the peak was fitted with a Gaussian plus a constant.  As lag
between the two bands we took the Gaussian center, rather than the DCF
maximum (see the arguments of Edelson et al. 1995 and 
Peterson et al. 1998). Table 1 reports
the lag and the 1$\sigma$ uncertainty (for all of the three sets of
observations).

\subsection{Modified Mean Deviation (MMD)}

The MMD technique was introduced by Hufnagel \& Bregman (1992). The
maximum correlation corresponds to a minimum of the MMD. The MMD and
relative Gaussian fit are given in Fig. 2 (MMD bin sizes are the same
of the DCF). The results are summarized in Table 1.

\subsection{Monte Carlo simulations}

In order to estimate the dependence of our findings on photon
statistics we followed the prescriptions of Peterson et al. (1998)
introducing "flux randomization" (FR) and "random subset selection"
(RSS). The DCF and MMD procedures were then applied to 5000 pairs of
Monte Carlo simulated light curves. The distribution of the lags are
reported in Fig. 2.  The mean lags and their uncertainties are obtained
by fitting these distributions with Gaussians (see Table 1).  For
comparisons, here we also include the results of this same analysis
for the 1997 $Beppo$SAX and 1994 ASCA data.

\begin{table}
\caption{Summary of the lags (hours)} \label{tbl-1}
\begin{center}\small
\begin{tabular}{|c|r|r|r|r|} \hline
&&&&\\
Obs. & DCF(1$\sigma$) & MMD(1$\sigma$) & FR/RSS DCF(1$\sigma$) &
FR/RSS MMD(1$\sigma$)\\ \hline
&&&&\\
SAX 1996 & 3.26(0.16) & 3.39(0.38) & 3.44(1.08) & 3.66(1.29)\\ 
SAX 1997 & 0.49(0.08) & 0.33(0.07) &0.37(0.47) & 0.32(0.33)\\ 
ASCA 1994 & 0.53(0.15) & 1.09(0.07) & 0.83(1.32) & 1.06(0.28)\\ 
&&&&\\
\hline
\end{tabular}
\end{center}
\end{table}

\section{Discussion}

It is apparent from Table 1 that the lags estimated with the two
techniques are fully compatible within the uncertainties of the Monte
Carlo results. The presence of a soft lag of 3 hours is rather clear in
the first $Beppo$SAX observation. The second $Beppo$SAX and ASCA
observations indicate a shorter lag which is consistent with
zero. Therefore the indication is that the lags are variable.
The variability of the lags is reminiscent of the variability between the
1991 and 1994 states of the source mentioned in the Introduction.

Time dependent models of BL Lac emission taking into account changes
in particle spectrum, energy losses, diffusion, and relativistic
propagation effects have been recently proposed (see Kirk et al. 1998;
Dermer 1998; Chiaberge \& Ghisellini 1998; Makino, this
conference). The soft lags appear qualitatively consistent with the
models.

\acknowledgments YHZ and AC acknowledge the Italian MURST for
financial support.


\begin{references}
\reference Chadwick, P. M., et al. 1998, ApJ, submitted
\reference Chiaberge, M., \& Ghisellini, G. 1998, M.N.R.A.S., submitted
\reference Chiappetti, L., et al. 1998, ApJ, to be submitted   
\reference Dermer, C. 1998, astro-ph/9805289
\reference Edelson, R. A., et al. 1995, ApJ, 438, 120
\reference Edelson, R. A., \& Krolik, J. H.  1988, ApJ, 333, 646
\reference Giommi, P., et al. 1998, A\&A, 333, L5 
\reference Kirk, J. G., et al. 1998, astro-ph/9801265
\reference Makino, F., et al. 1996 MPE Rep. 263, p. 413
\reference Maraschi, L., et al. 1998, astro-ph/9808177 
\reference Peterson, B. M., et al.,  1998, PASP, 110, 660
\reference Urry, C. M., et al. 1997, ApJ, 486, 799 
\end{references}
\end{document}